\documentclass[runningheads]{llncs}

\usepackage{amssymb}
\usepackage{graphicx}
\usepackage{epstopdf}
\usepackage{amsmath}
\usepackage{scalerel}
\usepackage{booktabs}
\usepackage{cases}
\usepackage{url}
\usepackage{array}
\usepackage{verbatimbox} 
\usepackage{color}
\usepackage{cite}

\newcommand{\nameWorkshop}{\textit{Mixed-Initiative ConveRsatiOnal Systems~}}
\newcommand{\acronymconf}{MICROS@ECIR2021}
\newcommand{\acronym}{MICROS~}

\urldef{\mailim}\path{ida.mele@iasi.cnr.it}
\urldef{\mailcm}\path{cristina.muntean@isti.cnr.it}
\urldef{\mailma}\path{m.aliannejadi@uva.nl}
\urldef{\mailnv}\path{n.voskarides@uva.nl}

\begin{document}

\mainmatter 

\title{MICROS: Mixed-Initiative {ConveRsatiOnal} Systems Workshop}

\author{Ida Mele\inst{1}\orcidID{0000-0002-3730-6383} \and Cristina Ioana Muntean\inst{2}\orcidID{0000-0001-5265-1831} \and \\ Mohammad Aliannejadi\inst{3}\orcidID{0000-0002-9447-4172}~\and~Nikos~Voskarides\inst{3}\orcidID{0000-0002-4850-6372}}

\institute{IASI-CNR, Rome, Italy \and ISTI-CNR, Pisa, Italy \and University of Amsterdam, Amsterdam, The Netherlands \\ \mailim ~~~ \mailcm \\ \mailma ~~~ \mailnv}

\authorrunning{Mele, Muntean, Aliannejadi, Voskarides}
\titlerunning{MICROS: Mixed-Initiative {ConveRsatiOnal} Systems Workshop}

\maketitle

\begin{abstract}
The 1st edition of the workshop on \nameWorkshop (\acronymconf) aims at investigating and collecting novel ideas and contributions in the field of conversational systems. Oftentimes, the users fulfill their information need using smartphones and home assistants. This has revolutionized the way users access online information, thus posing new challenges compared to  traditional search and recommendation.
The first edition of \acronym will have a particular focus on \textit{mixed-initiative conversational systems}. Indeed, conversational systems need to be proactive, proposing not only answers but also possible interpretations for ambiguous or vague requests. 
\end{abstract}


\section{Motivation and Topics of Interest}

The increasing popularity of personal assistant systems as well as smartphones has drawn attention to conversational systems with many application scenarios ranging from simple ones (e.g., checking the weather forecast) to more complex ones (e.g., performing e-commerce transactions).
Moreover, thanks to the recent advances in automatic speech recognition and voice generation, conversational assistants, such as Apple Siri or Microsoft Cortana, are widely being used in chatbots and smart-home devices  
as well as in wearable devices and smartphones.

Users employ conversational systems to seek information in an interactive way, often through voice interfaces. 
Information-seeking conversations can be categorized into two main classes: (i) search and (ii) recommendation. 
In \textit{conversational search}, answering users' requests poses several challenges.
First, the system must understand the user requests (a.k.a. questions, queries, or utterances) and return a ranked list of documents (results). The very top results (e.g., 1 or 2) must be potentially useful as the system replies vocally and thus it is impossible for the user to browse the list of results as in  Web search.
Another important challenge in conversational search is that (complex) information needs are not expressed with a single request; rather, the user formulates multiple subsequent questions that can be related to each other. In these multi-turn conversations, the current request may not be self-explanatory as the context is missing from the current question, but it was implied or explicitly mentioned in previous turns~\cite{voskarides-2020-query}. In particular, the subjects can be  pronouns referring to topics mentioned in the previous requests and/or answers~\cite{mele_SIGIR2020,voskarides-2020-query,AliannejadiCRC20}. Moreover, during the conversation, there might be slight or significant topic changes that need to be detected by the system~\cite{AliannejadiCRC20, mele_SIGIR2020,voskarides-2020-query}.
Furthermore, the users' requests  can be vague, ambiguous, or misleading.
Since the requests are formulated in natural language, they are prone to the ambiguity and polysemy of words, the presence of acronyms, mistakes, and grammar misuses. 
In such cases, the system can take the initiative by asking clarifying questions or by proposing keywords that disambiguate the request~\cite{AliannejadiZCC19,convai3}. 

In \textit{conversational recommendation}, the system interacts with the user asking for her opinion about some items~\cite{RadlinskiBBK19}. Preference elicitation introduces numerous challenges, such as modeling users' preference upon receiving their feedback and selecting the next question in a conversation to optimize the information gain. At the same time, the system should avoid any bias in the user's feedback.

We envision that advanced, flexible, and mixed-initiative interactions are very important in conversational systems as they allow the systems to identify the correct intent behind the user's requests and needs. 

\medskip
\noindent The workshop topics include but are not limited to:
\begin{enumerate}
    \item Applications of conversational search and recommendation systems
    \begin{itemize}
        \item Large-scale  retrieval candidate responses (e.g., documents, passages)  in conversational search
        \item Conversational and question-based recommendation systems
        \item Tracking information-need evolution during the conversation (e.g., context changes)
        \item Processing and rewriting of natural language conversational queries
        \item Relevance feedback in conversational search
    \end{itemize}
    
    \item Mixed-initiative interaction systems, such as clarification and preference elicitation in conversational systems
    \begin{itemize}
         \item Dialogue schema for conversational search
        \item Conversational navigation of search results
        \item Conversation history understanding and query modeling
        \item Pro-active search and recommendation interactions in conversational search
    \end{itemize}

    \item Deep learning and reinforcement learning for conversational search
    \begin{itemize}
        \item Conversational question answering
        \item Result summarization, explanation, and presentation in conversational search
        \item Balance and bias for more inclusive conversational systems
    \end{itemize}

    \item Multi-modal interactions for conversational interfaces (e.g., speech-only and small-screen interfaces)
    \begin{itemize}
        \item Voice-based search engine operations
        \item User intent and dialog state tracking in conversational search
        \item Personalization and user models for conversational search
    \end{itemize}

    \item Specialized applications and use cases for conversational search (e.g., health, finance, travel)

    \item Knowledge graph presentation in conversational search
    
    \item Data creation and curation for conversational search
   
    \item Evaluation metrics for effectiveness, engagement, satisfaction of conversational systems
\end{enumerate}

\section{Scientific Objectives}

The goal of the \acronym workshop is to collect and discuss novel approaches, evaluation techniques, datasets, and domain-specific applications of conversational systems. The workshop aims at bringing together academic and industry researchers
to create a forum for interacting and discussing the latest developments and new directions of research in the area of search- and recommendation-oriented conversational systems. 
These discussions are open to the whole audience and lead by experienced researchers from both academia and industry who actively participate in the workshop as keynote speakers and panelists.

A particular focus of  \acronym is on mixed-initiative interactions. This novel and still under-explored topic represents an important development in conversational systems. As a matter of fact, the interaction between the user and the system should go beyond the usual \textit{``user asks, system responds''} paradigm. Especially for those scenarios where the user requests are too generic, ambiguous, and may lack explicit subjects or context. The conversational system lacking enough confidence in identifying the topic of interest would take the initiative by asking the user to clarify her request, or proposing possible interpretations, or inferring the user's crisp opinion and interest.

\section{Organizing Team}

\begin{itemize}
\item \textbf{Ida Mele} is currently a researcher at IASI-CNR, Rome (Italy). She got her Ph.D. in Computer Engineering from Sapienza University of Rome. She has co-authored papers in peer-reviewed international conferences and top-tier journals. She has also served as PC member and reviewer for international conferences and journals. Her research interests are Web Mining, Information Retrieval, Recommendation Systems, and Social Media. Her current research focuses on conversational search and, in particular, on passage retrieval and re-ranking for multi-turn conversational searches.

    \item \textbf{Cristina Ioana Muntean} is a Researcher at ISTI-CNR, Pisa (Italy). Her main research interests are in Information Retrieval and Machine Learning with applications to Web search and social media. She is particularly interested in passage retrieval and conversational search using neural and classic IR models. She is an active member in the SIGIR, ECIR, CIKM, and TheWebConf communities, as author and part of the program committees.
    
    \item \textbf{Mohammad Aliannejadi} is a post-doctoral researcher at the University of Amsterdam (The Netherlands). His research interests include single- and mixed-initiative conversational information access and recommender systems. Previously, he completed his Ph.D.~at Universit\`a della Svizzera italiana (Switzerland), where he worked on novel approaches of information access in conversations. He has been an active member of the community, publishing and serving as a PC member in major venues and journals of the field. 
    
    \item \textbf{Nikos Voskarides} is a PhD candidate at the University of Amsterdam (The Netherlands). He is an active member of the community, publishing and serving as a PC member at major conferences such as SIGIR, ACL, EMNLP, ECIR and AKBC. His current research focuses on information retrieval for knowledge graphs and conversational search.

\end{itemize}

\bibliographystyle{splncs04}
\bibliography{refs} 

\end{document}